# Modeling Maintenance of Long-Term Potentiation in Clustered Synapses – Long-Term Memory Without Bistability

Paul Smolen

Department of Neurobiology and Anatomy
W. M. Keck Center for the Neurobiology of Learning and Memory
The University of Texas Medical School at Houston
E-mail: Paul.D.Smolen@uth.tmc.edu

**ABSTRACT**

Memories are stored, at least partly, as patterns of strong synapses. Given molecular turnover, how can synapses maintain strong for the years that memories can persist? Some models postulate that biochemical bistability maintains strong synapses. However, bistability should give a bimodal distribution of synaptic strength or weight, whereas current data show unimodal distributions for weights and for a correlated variable, dendritic spine volume. Thus it is important for models to simulate both unimodal distributions and long-term memory persistence. Here a model is developed that connects ongoing, competing processes of synaptic growth and weakening to stochastic processes of receptor insertion and removal in dendritic spines. The model simulates long-term (>1 yr) persistence of groups of strong synapses. A unimodal weight distribution results. For stability of this distribution it proved essential to incorporate resource competition between synapses organized into small clusters. With competition, these clusters are stable for years. These simulations concur with recent data to support the "clustered plasticity hypothesis" which suggests clusters, rather than single synaptic contacts, may be a fundamental unit for storage of long-term memory. The model makes empirical predictions, and may provide a framework to investigate mechanisms maintaining the balance between synaptic plasticity and stability of memory.

# INTRODUCTION

A central question in neuroscience is the mechanism by which memories can be preserved for years. Long-term memories are at least in part encoded as specific patterns, or "engrams", of strengthened synapses (Pastalkova et al., 2006; Whitlock et al., 2006), and long-term synaptic potentiation (LTP) persists for months *in vivo* (Abraham et al., 2002). How can specific groups of synapses remain strong for months or years despite turnover of macromolecules and fluctuations in the size and shape of synaptic structures?

Numerous mathematical models have been developed that hypothesize and describe maintenance of long-term memory (LTM) as dependent on bistability of synaptic weights, mediated by positive feedback loops of biochemical reactions, typically thought of as operative in dendritic spines. Proposed feedback mechanisms have relied on self-sustaining autophosphorylation of CaM kinase II (Lisman and Goldring, 1988; Miller et al., 2005), persistent phosphorylation of AMPA receptors by protein kinase A (Hayer and Bhalla, 2005), enhanced translation of protein kinase M $\zeta$ (Smolen et al., 2012), or self-sustaining clustering of a translation activator, cytoplasmic polyadenylation element binding protein (Si et al., 2010). With these models, LTP switches a synapse from a state of low basal weight to a high weight state, and also turns on the positive feedback loop. The loop then operates autonomously to keep the synapse in the high weight state indefinitely. However, despite extensive investigation, empirical evidence of a bistable distribution of two distinct synaptic weight states has not, in fact, been obtained. Although some studies (O'Connor et al., 2005; Petersen et al., 1998) have suggested two distinct strength states for synapses, as measured by the amplitude of excitatory postsynaptic currents before and after a stimulus protocol, these studies have only examined the early phase of LTP ($< 1$ h), which does not depend on protein synthesis or other processes necessary for long-term memory storage. Therefore those data do not address the dynamics of long-term memory storage. In addition, a demonstration of synaptic bistability would require not only finding two distinct synaptic strength states, but also finding that a set of different protocols for LTP induction (*e.g.,* different patterns of stimuli, or localized application of pharmacological agents) commonly switched synaptic weights between the <u>same</u> two stable states. Such a demonstration has not been attempted. In addition, modeling suggests that stochastic fluctuations of macromolecule numbers within a small volume such as a spine head are likely to destabilize steady states of biochemical positive feedback loops, causing randomly timed state switches (Bhalla, 2004; see Miller et al., 2005 for an opposing view). In hippocampal neuron cultures rapid, continuous, and extensive fluctuations of postsynaptic density (PSD) morphology are observed and are driven in part by synaptic activity (Blanpied et al., 2008). Such dynamics also seem difficult to reconcile with only two, or a few, stable well separated weight states.

Empirical distributions of the weights of excitatory synapses onto cortical or hippocampal pyramidal neurons appear unimodal (a single peak) rather than bimodal, and are commonly heavy-tailed (skewed towards high weights) (Barbour et al., 2007; Frick et al., 2007; Holmgren et al., 2003; Mason et al., 1991; Sayer et al., 1990; Song et al., 2005). Some empirical histograms are based on relatively small numbers of measurements, so that some degree of bimodality might be present but hidden in variability among bins. However, the weight distribution of Song et al. (2005) is based on measurements of several hundred excitatory postsynaptic potential (EPSP) amplitudes, and appears to particularly disfavor the bimodal hypothesis. A large number of measurements are fit well by a log-normal distribution (*i.e.*, a normal distribution with the logarithm of the volume on the x-axis).

In addition, a histogram of the volume of postsynaptic dendritic spines, based on a large number of individual measurements (~10,000) in mouse auditory cortex, is clearly unimodal, heavy-tailed, and approximately log-normal (Loewenstein et al., 2011). Observations support a substantial correlation between spine volume and synaptic weight. Spine volume is approximately proportional to the postsynaptic density size (Cane et al., 2014; Harris and Stevens, 1989; Katz et al., 2009), and to the number of postsynaptic AMPA receptors (Katz et al., 2009) as well as to the amplitude of the EPSC measured following localized glutamate uncaging (Matsuzaki et al., 2001).

If synaptic weights and correlated spine volumes are not in fact bistable, how can patterns of strong synapses be maintained for very long times? Two observations support a mechanism based on metastability of small clusters of large dendritic spines, corresponding to groups of strong synaptic contacts. The first observation is that although spine volumes fluctuate, some large spines are extremely stable, existing for months (in sensory cortex, Grutzlender et al., 2002; or in motor cortex, Yang et al., 2009). The second is that induction of late, protein synthesis-dependent LTP (L-LTP) at a spine facilitates L-LTP expression at nearby spines receiving stimuli too weak to support L-LTP if given alone (Govindarajan et al., 2011). This observation supports the "clustered plasticity hypothesis" in which clusters of spines on a single dendritic branch, rather than single spines, may serve as "primary functional units" for storage of LTM (Govindarajan et al., 2006). This hypothesis is now supported by substantial recent data (Winnubst and Lohmann, 2012). For example: 1) in motor cortex, learning induces coordinated formation of small spine clusters (Fu et al., 2012), 2) morphologically, spines are grouped into small clusters on pyramidal dendrites (Yadav et al., 2012), and 3) in rat hippocampal slice cultures, spontaneous co-activation of dendritic spines is frequent and is clustered, occurring more often for spines within 8 μm of each other (Takahashi et al., 2012).

Here an initial, relatively phenomenological model is developed describing synaptic weight changes due to competing processes of LTP (corresponding to spine growth) and long-term synaptic depression (LTD) (corresponding to spine shrinkage). Assuming that spine volume changes can be used as a proxy for weight changes, weight changes are simulated for discrete intervals of 1 day, over total times of months or years. This time interval was chosen to simulate the dynamics observed in experiments where volumes are imaged at intervals of ~ 1 day (Loewenstein et al., 2011; Yasumatsu et al., 2008). In the model, a synapse corresponds to a dendritic spine. Daily growth of synapses or spines corresponds to LTP, and daily shrinkage to LTD. Each day, the magnitude of LTP is given by a Gaussian random variable, as is that of LTD (Methods). These random variables are themselves proportional to the pre-existing weight, and this proportionality yields an approximately log-normal volume distribution at steady state. As suggested by some recent data (Grutzlender et al., 2002; Yang et al., 2009; Yasumatsu et al., 2008; but see Loewenstein et al., 2011), a volatility factor was introduced so that the weights of larger synapses fluctuate less. Such a factor is required in order that large synapses remain stable for simulated periods of months (see Discussion).

Model parameter sensitivity was lessened when synapses were modeled as coupled into small clusters (~10 spines). In accordance with data (De Roo and Muller, 2008; Yasumatsu et al., 2008), the model also incorporates disappearance or silencing of small synapses and compensatory regeneration of new synapses. With this model, when the dynamics of 1,000 small clusters were simulated, the weight distribution of the entire ensemble of individual synapses converged to a steady-state, log-normal form. Individual clusters remained stable for many years, with the average number of active synapses maintained in a range consistent with empirical data. The magnitude of daily changes in synaptic weight approximated a normal distribution but with an extra peak centered at $\Delta W = 0$, which constitutes a model prediction.

**METHODS**

Given the correlation between spine volume and synaptic weight, we refer to simulation of synaptic weights rather than spine volumes. Weight evolution in 1,000 independent clusters is simulated. Each cluster consists of $N_{cl}$ synapses, corresponding to $N_{cl}$ individual spines. At a given time most of these synapses are "active", with synaptic weight $W$ ranging from ~0.2 to 10. The remaining synapses are "silent", with a very low, basal weight of 0.05. Weight evolution is simulated using discrete, large time steps $\Delta t$, generally 1 day. At each time step all weights are synchronously updated. The size and direction of a weight update at a given synapse is assumed to be uncorrelated with that at a neighboring synapse, and also with the preceding update at the given synapse. These assumptions are supported by data describing spine volume changes on pyramidal dendrites (Yasumatsu et al., 2008).

Figure 1 schematizes key elements of the model. A cluster of four spines is illustrated, two are active. High values of $W$ correspond to large spines. Very small spines correspond to "silent" synapses. For each active synapse two independent processes change synaptic weights during each time step. An "LTP" process increases $W$ and an "LTD" process decreases $W$. Strong synapses consume more resources for their maintenance, corresponding to locally available mRNAs/proteins. The model thus assumes that the more strong synapses present in a cluster, the fewer resources are available to support synaptic growth (LTP). Thus the amplitude of LTP decreases with the number of strong spines. Empirically, smaller spines are more volatile. Thus the model assumes that the amplitude of LTP is also proportional to a volatility factor that decreases as $W$ increases (Eqs. 1-2), The amplitude of LTD is also proportional to this factor (Eq. 3). LTP and LTD add together to give the net change in $W$ per time step (Eq. 4).

Small spines frequently disappear (Yasumatsu et al., 2008). To balance disappearance new spines form, preferentially close to active synapses, maintaining clusters of active synapses (De Roo and Muller, 2008). To represent these dynamics the model allows regeneration of synapses, within each cluster, only if a synapse adjacent to the one being regenerated is strong (Eq. 5).

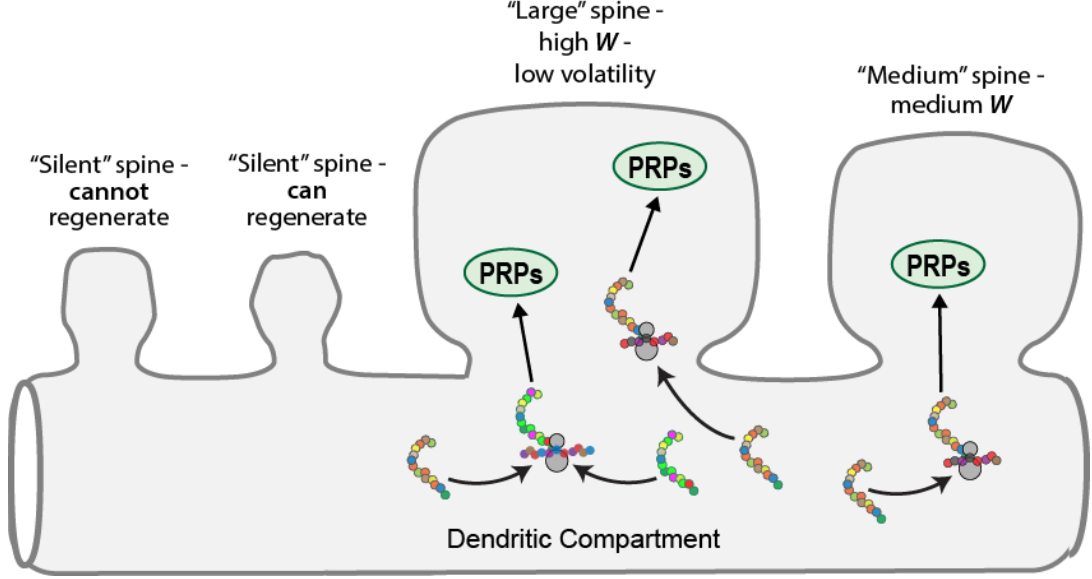

**FIGURE 1**. Schematic illustrating key model elements and assumptions. Two silent synapses, corresponding to two small dendritic spines, are adjacent to two active synapses corresponding to larger spines. Only the silent synapse next to an active synapse is currently eligible for regeneration to an active state. Although the dynamics of mRNAs or of locally synthesized proteins are not simulated, large synapses are implicitly assumed to capture more of these resources. This is illustrated by a greater flow of mRNAs (colored strands) onto ribosomes near the largest synapse, and a greater flow of plasticity-related proteins (PRPs) into that synapse. This resource capture is assumed to attenuate the amount of resource available to support growth of all synapses in the cluster. The average amplitude of LTP events, responsible for positive increments in synaptic weight $W$ at each time step, is thereby reduced when more synapses in a given cluster are strong.

For each time step, the LTP and LTD amplitudes are proportional, respectively, to Gaussian random variables $r_1$ and $r_2$. These variables have respective means $a_1$ and $a_2$, and standard deviations $sd_1$ and $sd_2$. $a_1$ and $a_2$ are substantially larger than $sd_1$ and $sd_2$ so that $r_1$ and $r_2$ are very rarely negative, but if either $r_1$ or $r_2$ becomes negative it is reset to 0 so that LTP and LTD amplitudes are always non-negative. For the parameter $a_2$, $sd_2 = 0.25 a_2$. A synapse is "strong" if its weight is above a threshold $T_{st}$. The average LTP amplitude $a_1$ is a decreasing function of the number of strong synapses in a given cluster, denoted $N_{st}$. With $N_{cl}$ the total number of synapses in a cluster, the average LTP amplitude $a_1$ decreases linearly with $N_{st}$, from a maximum amplitude $x_2$ (for $N_{st} = 0$) to a minimum $x_1$ (for $N_{st} = N_{cl}$). $sd_1 = 0.25 a_1$.

LTP and LTD are proportional to a volatility factor $VO_W$, decreasing with $W$:

$$VO_W = \left\{ v_{hi} - (v_{hi} - v_{lo}) \left( \frac{W}{W + W_{med}} \right) \right\} \qquad (1)$$

From Eq. 1, $VO_W$ decreases from the parameter $v_{hi}$ (for $W = 0$) to $v_{lo}$ (for $W \gg W_{med}$). When $W = W_{med}$, $VO_W$ is midway between $v_{hi}$ and $v_{lo}$.

LTP and LTD amplitudes are also proportional to the pre-existing value of $W$. To keep W bounded the LTP amplitude is also multiplied by a decreasing function of W that has two parameters, $k_{hi}$ and $W_{hi}$. Combining factors the LTP amplitude is:

$$A_{LTP} = W \cdot r_1 \cdot VO_W \left[ 1 - k_{hi} \left( \frac{W}{W + W_{hi}} \right) \right] \quad (2)$$

The LTD amplitude is:

$$A_{LTD} = W \cdot r_2 \cdot VO_W \quad (3)$$

At each time step, for each active synapse:

$$W_{new} = W_{old} + A_{LTP} - A_{LTD} \quad (4)$$

If $W$ falls below a threshold $T_{wk}$, the synapse is reset to be silent. For each silent synapse at each time step, the probability for regeneration $P_{ACT}$ increases with the number of strong synapses in its cluster, to a maximal value $P_{bas}$:

$$P_{ACT} = P_{bas} \frac{N_{st}}{N_{cl}} \quad (5)$$

Also, regeneration only occurs if an adjacent synapse is strong. In a cluster synapse 1 can only switch to active if synapse 2 is strong, and synapse 5 can only switch if synapse 4 and/or 6 is strong. A switch resets $W$ to $W_{reset}$, above $T_{wk}$.

For all simulations, to ensure initial weight distributions were at equilibrium, illustrated distributions, responses to stimuli, and other quantities were computed only after 50,000 simulated days. $W$ is non-dimensional and $t$ has units of hrs.

Simulation of the Pearson correlation coefficient R of synaptic weights was done as follows. Let $X_i$ denote the total set of $n$ synaptic weights at a reference time, with i the indexing variable from 1 to $n$. For example, for 1,000 10-synapse clusters, $n = 10,000$. Let $Y_i$ denote the set of $n$ weights at any later time step. As $t$ increases from the reference time, $Y_i$ will evolve, and R will decline from 1. Let $\bar{X}$, $\bar{Y}$ denote the means of $X_i$, $Y_i$. The standard equation to calculate R was used:

$$R = \frac{\sum_{i=1}^{n}\left[(X_i - \bar{X})(Y_i - \bar{Y})\right]}{\sqrt{\sum_{i=1}^{n}(X_i - \bar{X})^2}\sqrt{\sum_{i=1}^{n}(Y_i - \bar{Y})^2}} \tag{6}$$

Standard model parameter values, used in all simulations unless stated otherwise, are as follows:

$N_{cl} = 10$, $W_{reset} = 0.4$, $T_{wk} = 0.08$, $T_{st} = 0.8$, $v_{hi} = 4.0$, $v_{lo} = 0.2$, $W_{med} = 0.4$, $x_1 = 0.144$, $x_2 = 0.18$, $a_2 = 0.16$, $k_{hi} = 0.05$, $W_{hi} = 20.0$, $P_{bas} = 0.1$.

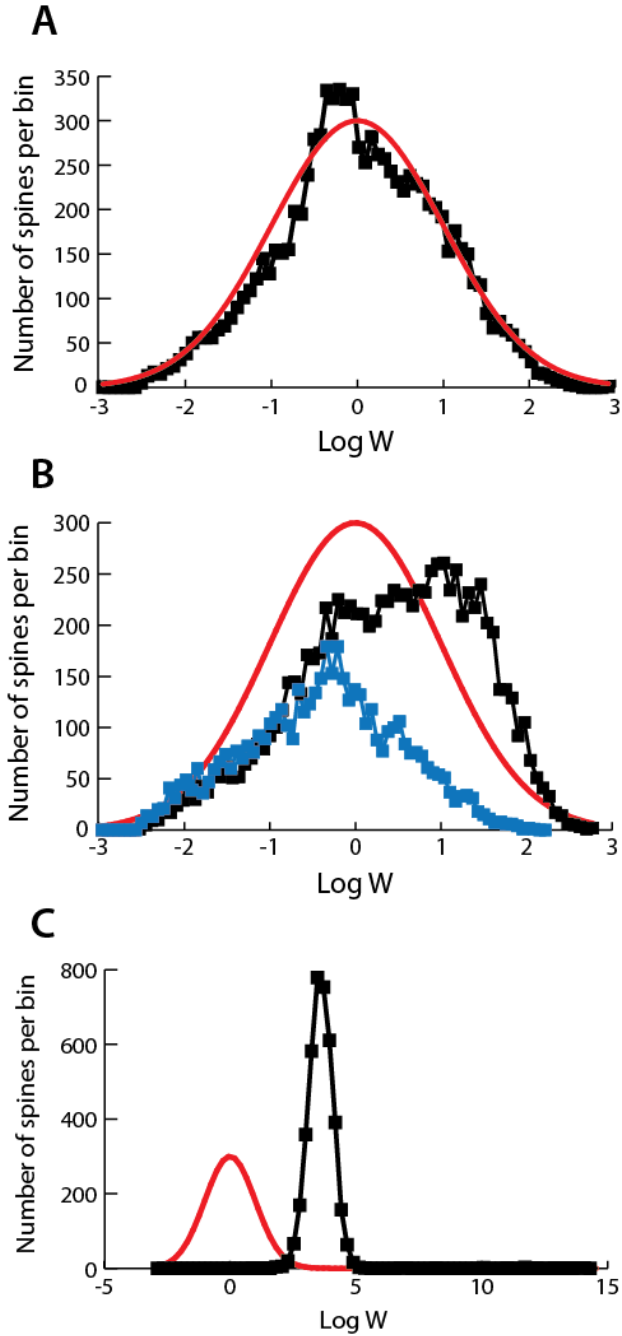

## RESULTS

Figure 2A illustrates the approximately log-normal distribution of synaptic weights obtained at steady state. Here, 1,000 clusters of synapses were simulated, with $N_{cl} = 10$. The black trace is the resulting histogram of synaptic weight $W$, the red trace is a log-normal distribution that approximates the histogram. There are 10,000 values of $W$ in the histogram. The range of $W$ spans approximately 5 natural log units (i.e., a multiplicative factor of ~150). This range is similar to data describing the range of dendritic spine volumes (Loewenstein et al., 2011; Yasumatsu et al., 2008).

FIGURE 2  Simulated distributions of synaptic weights. **(A)** Distribution for 1,000 independent clusters with $N_{cl} = 10$. Black trace, histogram with 80 bins illustrating an approximately log-normal distribution of the 10,000 weights. Each bin is equal in width in natural log units. Red curve, a log-normal distribution (mean at 0.0, standard deviation of 1.0), fitted by trial and error, that approximately reproduces the histogram. Standard parameter values were used. The histogram was constructed after 50,000 simulated days to ensure a steady state. **(B)** Black and red traces, similar to A, except the mean and standard deviation of LTP, parameters $a_1$ and $sd_1$, are fixed at 0.16 and 0.04 respectively. Blue trace, the histogram of $W$ is shifted to much lower values when $a_1$ is decreased by 2%. **(C)** Weight dynamics without regeneration of silent synapses. The histogram of the weights of active synapses shifts to much greater values (black trace). The red curve shows the same normal distribution as in **A**. Approximately 45% of synapses are silent and unable to regenerate, and are not included in the histogram. The histogram was constructed after 50,000 simulated days. It is an approximate steady state, although with no regeneration, all synapses would become silent after a much longer time.

To prevent this distribution from being overly sensitive to the average daily LTP amplitude, the model assumes this amplitude decreases with the number of strong synapses in the cluster to which the synapse belongs. With this amplitude decrease, starting from Fig. 2A, when the mean LTP amplitude was decreased by 5%, the mean of W decreased by only 16%. In contrast, Figure 2B illustrates that in a variant without this amplitude decrease, with mean and standard deviation of the LTP amplitude fixed, the weight histogram was shifted to the right of the red log-normal distribution (from Fig. 2A), and has a shape clearly distorted from a log-normal distribution, with a much steeper cutoff at high W. To attempt to improve the weight histogram, the mean LTP amplitude was decreased by 2%. This small change resulted in a large shift of the histogram to the left (blue trace), and 56% of the synapses became silent (not included in the histogram). This model variant was not analyzed further due to its extreme sensitivity. Following results are based on the simulation of Fig. 2A with standard parameter values (Methods).

Removal of synaptic regeneration greatly alters the distribution and dynamics of synaptic weights. The distribution (Fig. 2C) no longer resembles empirical data. The distribution of weights is strongly bimodal, with almost 50% of synapses silenced at the low basal weight, unable to regenerate, and the remainder in a narrow distribution centered at very high weights.

Two typical time courses for clusters with regeneration are illustrated in Figs. 3A and 3B.

Figure 3A-B illustrates that strong synapses are much more stable than weak synapses on average, in agreement with data demonstrating that larger dendritic spines are more persistent (Holtmaat et al., 2005; Trachtenberg et al., 2002; Yasumatsu et al., 2008). Strong synapses often maintain high values of W for a year or more, whereas weak synapses show much larger relative (fractional) fluctuations in W. Weak synapses often drop to a very low basal weight. These "silent" synapses can regenerate, with their weights reset to $W_{reset} = 0.4$. Regeneration is evident as vertical jumps in time courses near the bottom of Figs. 3A-B (*e.g.* arrowheads below *x* axes). Figure 3C illustrates a typical time course for the number of strong synapses $N_{st}$ in a cluster (with weights greater than the threshold $T_{st} = 0.8$). Clusters are very stable in that, for $N_{cl} = 10$, $N_{st}$ almost always remains between 4 and 7 for years.

Minerbi et al. (2009) recorded synaptic dynamics for many days in cortical neuron cultures. They examined variations in the size of PSD-95:GFP puncta. Their dynamics are qualitatively consistent with Figs. 2A and 3 in that: 1) new synapses were continually formed, 2) synapses whose size was reduced beneath some threshold, analogous to the model threshold $T_{wk}$, were eliminated, and 3) large synapses tended to decrease in size and small synapses to increase in size. In accordance with 3), stability of a simulated steady-state, unimodal distribution such as that of Fig. 2A requires that individual synapses with weights above the peak of the distribution decrease their weight on average, and *vice versa* for those with weights below the peak. Empirically, Matsuzaki et al. (2004) found that smaller spines are indeed more likely to undergo stable enlargement in response to an LTP induction protocol.

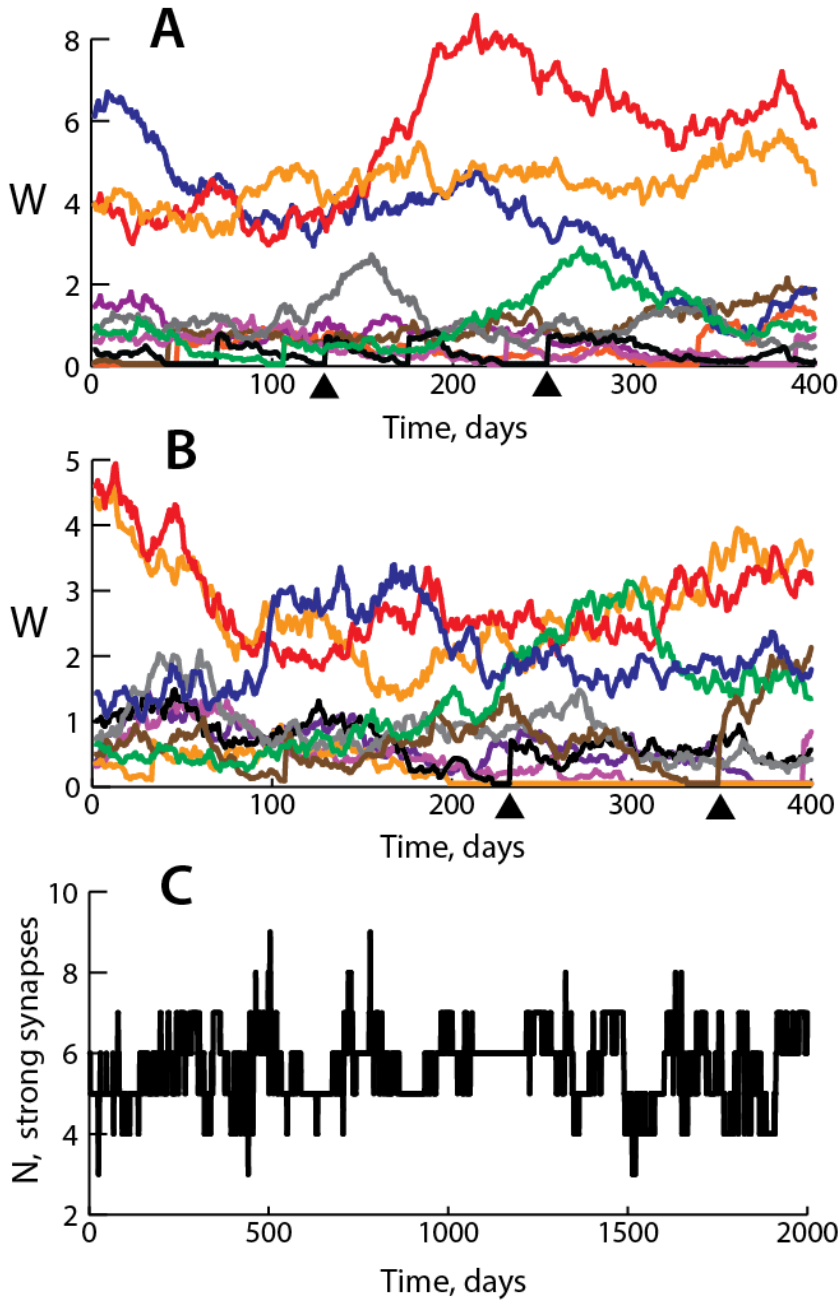

FIGURE 3  Dynamics of synaptic weights. **(A) (B)** Representative time courses of the synaptic weights in 10-synapse clusters, over 400 days. For each cluster, synapses with higher average $W$ exhibit less volatility (smaller percent changes in $W$). Synapses with high average weights often remain strong for periods > 1 yr. **(C)** A representative time course of the number of strong synapses, $N_{st}$, in a 10-synapse cluster. Although individual weights undergo large fluctuations, $N_{st}$ is extremely stable, remaining between 4 and 7 for periods of over a year, and rarely leaving this range during 5.5 years.

Loewenstein et al. (2011) and Yasumatsu et al. (2008) both present plots of daily changes in spine volume vs. initial spine volume. For comparison, Figure 4 illustrates simulated distributions of the amplitudes of the daily changes in synaptic weight, $\Delta W$, at steady state.

The majority of the histogram of Fig. 4A is seen to be fitted well by a normal distribution, with the clear exception of the narrow peak close to $\Delta W = 0$. A scatter plot of $\Delta W$ *vs.* $W$ (not shown) revealed that over almost the entire range of $W$, the amplitude of $\Delta W$ varied from near zero to a peak value of ~ 0.3 – 0.5. Therefore the narrow peak near zero is not due exclusively to large synapses.

Figure 4B plots daily changes in weight vs. initial weight, and illustrates that the average of the change in $W$ varies much less than does $W$ itself. As $W$ increases from 0.05 to 2.0, the absolute value of $\Delta W$ increases only from ~0.05 to 0.12. Thus the <u>relative</u> change in $W$ (*i.e.* $\Delta W / W$), decreases substantially as $W$ increases. This prediction of the model appears in accordance with the data from Yasumatsu et al. (2008), but not with the data of Loewenstein et al. (2011) for which this relative change appears nearly constant, see Discussion.

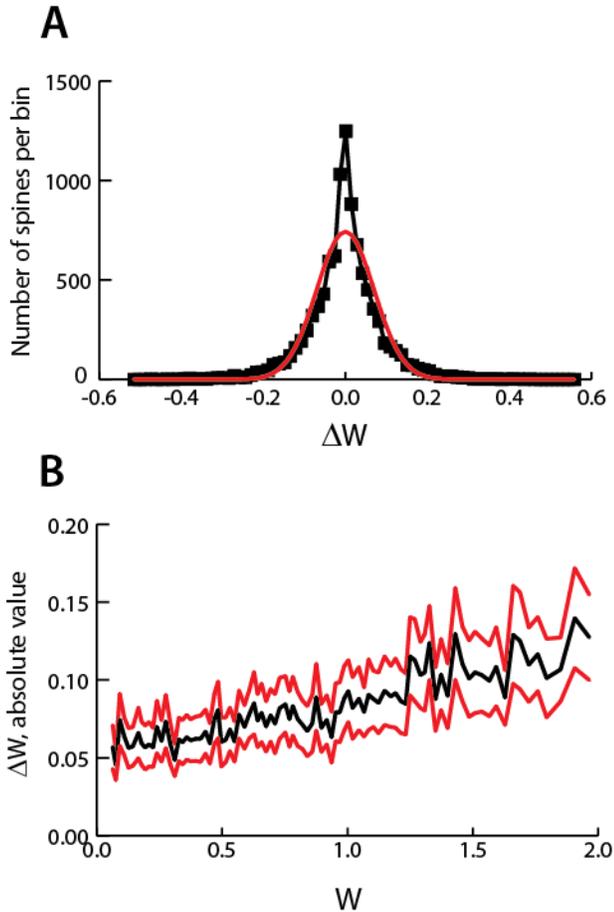

FIGURE 4 Distributions of the magnitude of synaptic weight change. **(A)** Black trace, histogram with 80 bins illustrating the steady-state distribution of the weight update amplitudes for the active synapses in the simulation of Fig. 2A ($\Delta W = W_{new} - W_{old}$, Eq. 4 in Methods). These amplitudes consist of all the synchronous weight updates for the 9,482 active synapses (out of 10,000 total), at the time step immediately after 50,000 simulated days. Red curve, a normal distribution (mean 0.0, standard deviation 0.07) that approximates the histogram excepting the sharp peak for update amplitudes ($\Delta W$) near zero. **(B)** A histogram of $\Delta W$ vs. $W$ more clearly shows a slightly increasing, relatively linear trend. 80 bins for $W$ were equally spaced on a log scale. Black trace, mean values of $\Delta W$. Red traces, mean ±1 standard deviation.

The above simulations illustrate that individual strong synapses can maintain high weights for many months. However, what if the weight distribution is altered by an imposed, large perturbation, that sets high weights for a specified subset of synapses? Such a perturbation may correspond to formation of a specific, long-term memory engram. In the model, will such a group of strong synaptic weights remain elevated for months, corresponding to long-lasting storage of a specific memory? Starting from the steady-state weight distribution of Fig. 2A, for all of the 1000 10-synapse clusters, synapses 1-5 were reset to a high weight of 5.0 at $t = 200$ days. This weight is well above the steady-state mean $W$. The other 5 synapses were reset to a low weight (0.5). We then simulated the dynamics of all clusters for a further 700 days. Figure 5A below illustrates a typical time course of weights for a cluster. After the reset, the weights fluctuate but 400 days later, all but one of the strong synapses have maintained $W$ above the steady-state average. Figure 5B illustrates the time course of resetting and decay for all 5,000 synapses that were reset to $W = 5.0$. Their average weight decays slowly, such that 700 days after reset, this average is still about twice the steady-state average of $W$. The lower red trace, one standard deviation below the average weight, is also still above the steady-state average.

If the steady-state distribution of $W$ is evolved without any perturbation, the time scale for decorrelation of synaptic weights from their specific values at a given time is, perhaps surprisingly, quite long, similar to the time scale for the decay of perturbed synaptic weights. Starting from that steady-state distribution, the Pearson correlation coefficient between the weights of the 10,000 synapses decays with a time constant of approximately 500 days (Fig. 5C), similar to the dynamics of Fig. 5B.

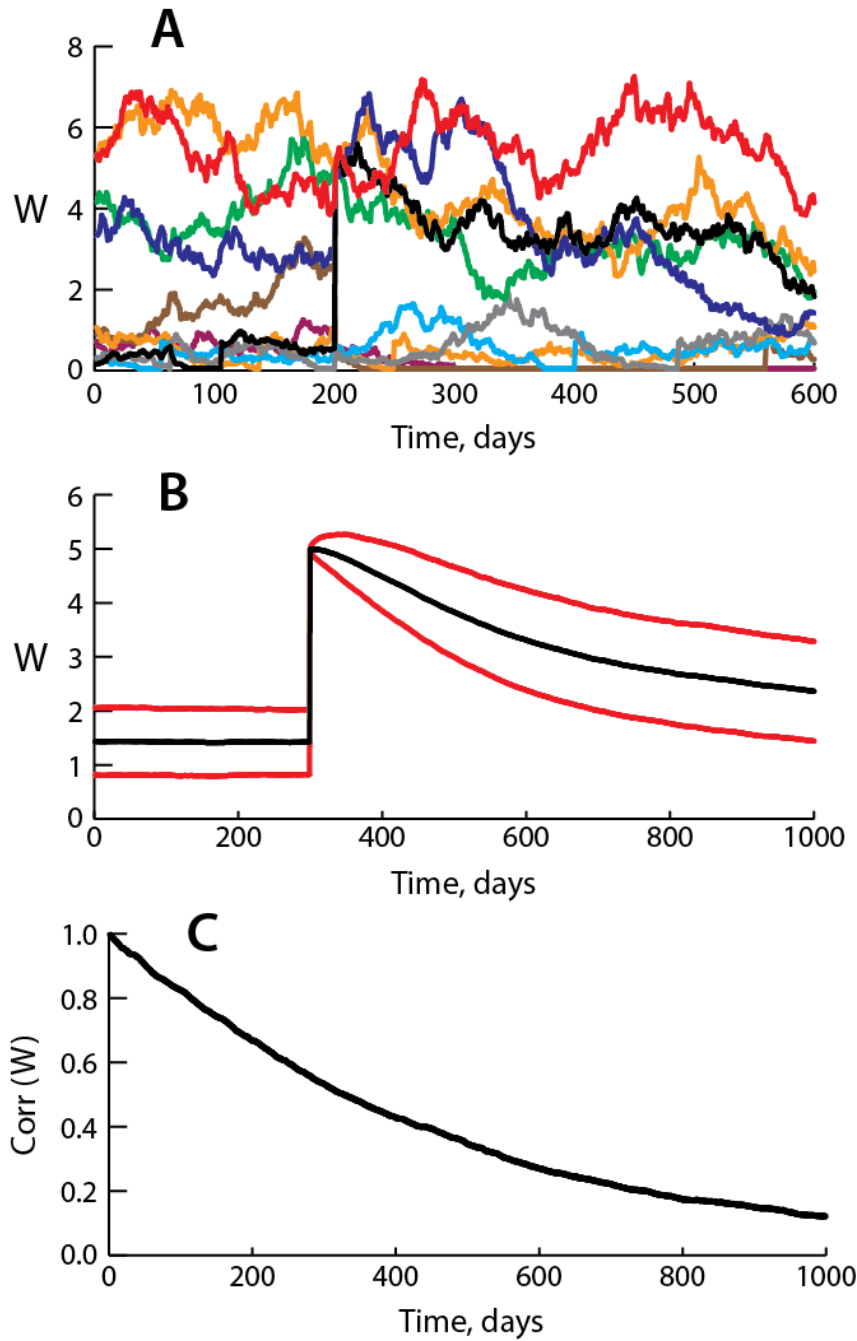

FIGURE 5  Weight dynamics with imposed LTP of a subset of synapses. (**A**) A representative time course of a 10-synapse cluster. $t = 0$ corresponds to the distribution of Fig. 2A. At $t = 200$ days, W was set to a high value of 5 for synapses 1 – 5, and a low value of 0.5 for synapses 6 – 10. Over 400 days, the distinction between strong and weak synapses was largely preserved, with synapses 1 – 4 remaining at higher weights than synapses 5 – 10. (**B**) Dynamics of the potentiated synapses in 10,000 10-synapse clusters. For each cluster, synapses 1 – 5 were potentiated as in **A**, thus the dynamics of 5000 synapses are illustrated. Black trace, time course of the average weight of these synapses. Red traces, ± 1 standard deviation from average. 700 days after LTP, the average weight remains > 1 standard deviation above the steady-state average. (**C**) Correlation coefficient describing the evolution of synaptic weights during maintenance of the steady-state distribution. For 10,000 synapses in 10-synapse clusters, starting from the distribution of Fig. 2A, the Pearson correlation coefficient R was calculated between the array of values of W at $t = 0$ and the array at later times.

## DISCUSSION

With simulations of isolated, uncoupled synapses (Fig. 2B), weight distributions were extremely sensitive to model parameters. This model variant was therefore not a plausible description of synaptic dynamics, because biophysical parameters are expected to vary somewhat between dendritic spines, dendritic branches, and individual neurons. This parameter sensitivity was eliminated when synapses were modeled as coupled into small clusters (~10 synapses). With this coupling, when the dynamics of 1,000 clusters were simulated, the weight distribution of individual synapses converged to a stable steady-state, log-normal form (Fig. 2A). Individual synaptic weights continue to fluctuate but this weight distribution is stable indefinitely.

To simulate a distribution spanning a broad range of synaptic weights, as has been observed empirically, it was also critical to model synaptic regeneration. To obtain distributions similar to that of Fig. 2A, synapses that fell to a low basal weight needed to have, on successive days, a probability of weight reset to a higher, "active" value. This method of simulating synaptic regeneration was chosen in order to maintain equal average numbers of synaptic loss and regeneration events within any given cluster, so that the mean number of active synapses in a cluster remains stable over time. Empirically, the number of active spines in a cluster is relatively low (~3-7) (Yadav et al., 2012). Thus for clusters with low numbers of active spines to serve as functional memory storage units, as suggested by the clustered plasticity hypothesis (Govindarajan et al., 2006), the average cluster size would need to be stable to avoid disappearance of clusters.

Within a cluster, empirical data illustrates that spines compete for LTP expression, *i.e.,* local resources (possibly amounts of key proteins) are limited such that the amount of LTP at a given spine decreases if LTP is simultaneously induced at multiple spines in the same cluster (Govindarajan et al., 2011). Our mechanism of coupling synapses into clusters is similar in that it corresponds to an additional form of resource competition between synapses, in which competition is generated by ongoing maintenance of multiple synapses rather than only by simultaneous LTP of multiple synapses. Specifically, the mean magnitude of LTP at a given synapse during a simulated day was assumed to be a decreasing function of the number of other large synapses in the same cluster (Methods). Thus, ongoing maintenance of large spines was assumed to consume resources (proteins, RNA) that would otherwise be available for strengthening of neighboring spines, so that their mean LTP magnitude decreased. Current data does not support or refute this specific coupling mechanism. However, it appears plausible, and it constitutes a model prediction.

The model of Loewenstein et al. (2011) simulates daily changes in the volume of dendritic spines, and obtains a steady-state log-normal distribution of spine volumes very similar to the empirical distribution found by these authors. These important results have clarified the necessity of reconciling unimodal weight distributions with stable storage of long-term memory. Their model consists of an Ohrnstein-Uhlenbeck stochastic process in which the logarithm of the volume of any given spine is directly incremented each day. The model presented here may constitute a significant further advance, in that it represents more biophysical elements, such as synapse loss / regeneration and synapse clustering. The LTP and LTD processes in the model: 1) act directly on the synaptic weight rather than on its logarithm, and 2) can be thought of as due to a large number of individual biochemical events occurring during a simulated day, and corresponding to insertion and removal of individual molecular complexes or slots. Point 2) connects the current model with the more detailed molecular model of Lisman and Raghavachari (2006), in which LTP and LTD correspond respectively to insertion and removal of "slot" complexes, each consisting of a small number of AMPA receptors together with associated scaffolding proteins and, possibly, kinases or other signaling proteins.

Recent data (Loebel et al., 2013) are consistent with these ideas and suggest such slot complexes include presynaptic components. The following considerations suggest that the dynamics of the model presented in this manuscript are consistent with the model of Lisman and Raghavachari (2006). If over the course of a day, the time intervals between individual insertions of "slot" complexes as well as the intervals between individual removals of complexes each follow Poisson distributions with a relatively large (> 20) mean number of individual events, or follow Gaussian distributions, then the sum of each of these Poisson (or Gaussian) processes will closely approximate a Gaussian random variable. These Gaussian variables would correspond to the random variables $r_1$ and $r_2$ that the total daily LTP and LTD amplitudes are proportional to. It also appears plausible that average numbers of slot insertions and removals are approximately proportional to the pre-existing size of a dendritic spine, corresponding to the model assumption that mean LTP and LTD amplitudes are proportional to synaptic weight. Thus, although the present model does not yet represent dynamics of specific molecular species, predicted dynamics appear consistent with postulated molecular processes.

In the model, the average relative change in $W$ over a simulated day, $\Delta W / W$, decreases substantially as $W$ increases (Fig. 4B). This result is essential for the model to represent stable long-term memory storage, because selective stability of the weight of strong synapses is only found if the relative change in $W$ decreases in this manner. Are these dynamics supported by current data? Relevant empirical data describes the relationship of spine volume changes $\Delta V$ to $V$. These data are, however, contradictory. Loewenstein et al. (2011) illustrate a substantially larger relative variation in $\Delta V$ (their Fig. 4C), such that $\Delta V$ and $V$ appear approximately proportional. However, Yasumatsu et al. (2008) show a different relationship, for which smaller spines have an average $\Delta V$ similar to large spines (their Fig. 1C). The latter relationship, but not the former, appears compatible with our assumption. Therefore further empirical investigation is needed to clarify this critical aspect of synaptic dynamics.

For a given day at steady state, the magnitudes of the daily changes in weight were distributed approximately normally except for an extra peak centered at $\Delta V = 0$, which constitutes another model prediction (Fig. 4A). Empirical data has not reported such a peak close to zero (Loewenstein et al. 2011). However, data that a minor percentage of spines are stable for months or years (Grutzlender et al., 2002; Yang et al., 2009) suggests such a peak might exist but not yet be resolved due to empirical sensitivity limits. The current model does not fully represent the biochemical processes that may underlie long-term stability of such a subset of spines. However, recent studies support the hypothesis that ongoing, spontaneous neuronal activity is critical for long-term maintenance of synaptic strength. Models have demonstrated that such activity could preferentially maintain synapses that are already strong (Tetzlaff et al., 2013) possibly by preferentially re-activating stored patterns of strengthened synapses (Wei and Koulakov, 2014). Experiment has demonstrated that inducibly and reversibly knocking out NMDA receptor function can irreversibly eliminate remote memories (Cui et al., 2004), plausibly by preventing spontaneous activity from potentiating and thereby maintaining synapses.

The simulations of Fig. 5A-B illustrate that the model can store a specific memory trace, for a period on the order of a year, corresponding to persistence of a pattern of strong synapses that could serve as the engram for a long-term memory. However in humans, some memories persist for many years. In accordance with the hypothesis of Govindarajan et al. (2006), such memories might be encoded at the cluster level rather than the single-synapse level, as a set of specific, stable clusters of active synapses. In simulations, these clusters were stable indefinitely (Fig. 3C). They maintained a range of active synapse numbers (~4-7) similar to the range suggested by data demonstrating clustering on neocortical pyramidal dendrites (Yadav et al., 2012). Because each cluster was stable indefinitely, this simulation suggests that if a pattern of such clusters was established, that pattern could persist for years. The ongoing, daily LTP increments that are a necessary component of the model may correspond, at least in part, to frequent empirical re-strengthening of synapses by ongoing spontaneous (or environmentally induced) activity, which may induce repeated cycles of NMDA receptor-dependent LTP at synapses in established memory engrams.

The model makes additional predictions. When comparing spines of similar sizes in different clusters, the average volume change between imaging sessions is expected to be less positive (or more negative) if other spines in a cluster are large. And, because synaptic weights and spine volumes are not considered bistable, different induction protocols for late, protein-synthesis dependent LTP are predicted to induce clearly different amplitudes of synaptic weight increase, or of average spine volume increase. Furthermore, without bistability, repeated applications of stimulus protocols should, at least in some cases, further enhance L-LTP.

**ACKNOWLEDGEMENTS**

Supported by NIH grant R01 NS073974.

# REFERENCES

Abraham WC, Logan B, Greenwood JM, Dragunow M (2002). Induction and experience-dependent consolidation of stable long-term potentiation lasting months in the hippocampus. J. Neurosci. 22: 9626-9634.

Barbour B, Brunel N, Hakim V, Nadal JP (2007) What can we learn from synaptic weight distributions? Trends Neurosci. 30: 622-629.

Bhalla US (2004) Signaling in small subcellular volumes. II. Stochastic and diffusion effects on synaptic network properties. Biophys. J. 87: 745-753.

Blanpied TA, Kerr JM, Ehlers MD (2008) Structural plasticity with preserved topology in the postsynaptic protein network. Proc. Natl. Acad. Sci. USA 105: 12587-12592.

Cane M, Maco B, Knott G, Holtmaat A (2014) The relationship between PSD-95 clustering and spine stability *in vivo*. J Neurosci. 34: 2075-2086.

Cui Z, Wang H, Tan Y, Zaia KA, Zhang S, Tsien JZ (2004) Inducible and reversible NR1 knockout reveals crucial role of the NMDA receptor in preserving remote memories in the brain. Neuron 41: 781-793.

De Roo M, Klauser P, Muller D (2008) LTP promotes a selective long-term stabilization and clustering of dendritic spines. PLoS Biol. 6: e219.

Frick A, Feldmeyer D, Helmstaedter M, Sakmann B (2008) Monosynaptic connections between pairs of 15a pyramidal neurons in columns of juvenile rat somatosensory cortex. Cereb. Cortex 18: 397-406.

Fu M, Yu X, Lu J, Zuo Y (2012) Repetitive motor learning induces coordinated formation of clustered dendritic spines *in vivo*. Nature 483: 92-96.

Govindarajan A, Kelleher RJ, Tonegawa S (2006) A clustered plasticity model of long-term memory engrams. Nat. Rev. Neurosci. 7: 575-583.

Govindarajan A, Israely I, Huang SY, Tonegawa S (2011) The dendritic branch is the preferred integrative unit for protein synthesis-dependent LTP. Neuron 69: 132-146.

Grutzendler J, Kasthuri N, Gan WB (2002) Long-term dendritic spine stability in the adult cortex. Nature 420: 812-816.

Harris KM, Stevens JK (1989) Dendritic spines of CA1 pyramidal cells in the rat hippocampus: serial electron microscopy with reference to their biophysical characteristics. J. Neurosci. 9: 2982-2997.

Hayer A, Bhalla US (2005) Molecular switches at the synapse emerge from receptor and kinase traffic. PLoS Comput. Biol. 1: 137-154.

Holmgren C, Harkany T, Svennenfors B, Zilberter Y (2003) Pyramidal cell communication within local networks in layer 2/3 of rat neocortex. J. Physiol. 551: 139-153.

Holtmaat AJ, Trachtenberg JT, Wilbrecht L, Shepherd GM, Zhang X, Knott GW, Svoboda K (2005) Transient and persistent dendritic spines in the neocortex in vivo. Neuron 45: 279-291.

Katz Y, Menon V, Nicholson DA, Geinisman Y, Kath WL, Spruston N (2009) Synapse distribution suggests a two-stage model of dendritic integration in CA1 pyramidal neurons. Neuron 63: 171-177.

Lisman JE, Goldring MA (1988) Feasibility of long-term storage of graded information by the $Ca^{2+}$ / calmodulin-dependent protein kinase molecules of the postsynaptic density. Proc. Natl. Acad. Sci. USA 85: 5320-5324.

Lisman J, Raghavachari S (2006) A unified model of the presynaptic and postsynaptic changes during LTP at CA1 synapses. Sci. STKE 2006: re11.

Loebel A, Le Be JV, Richardson MJ, Markram H, Herz AV (2013) Matched pre- and post-synaptic changes underlie synaptic plasticity over long time scales. J. Neurosci. 33: 6257-6266.

Loewenstein Y, Kuras A, Rumpel S (2011) Multiplicative dynamics underlie the emergence of the log-normal distribution of spine sizes in the neocortex *in vivo*. J. Neurosci. 31: 9481-9488.

Mason A, Nicoll A, Stratford K (1991) Synaptic transmission between individual pyramidal neurons of the rat visual cortex in vitro. J. Neurosci. 11: 72-84.

Matsuzaki M, Ellis-Davies GC, Nemoto T, Miyashita Y, Iino M, Kasai H (2001) Dendritic spine geometry is critical for AMPA receptor expression in hippocampal CA1 pyramidal neurons. Nat. Neurosci. 4: 1086-1092.

Matsuzaki M, Honkura N, Ellis-Davies GC, Kasai H (2004) Structural basis of long-term potentiation in single dendritic spines. Nature 429: 761-766.

Miller P, Zhabotinsky AM, Lisman JE, Wang XJ (2005) The stability of a stochastic CaMKII switch: dependence on the number of enzyme molecules and protein turnover. PLoS Biol. 3: e107.

Minerbi A, Kahana R, Goldfeld L, Kaufman M, Marom S, Ziv NE (2009) Long-term relationships between synaptic tenacity, synaptic remodeling, and network activity. PLoS Biol. 7: e1000136.

O'Connor DH, Wittenberg GM, Wang SS (2005) Graded bidirectional synaptic plasticity is comprised of switch-like unitary events. Proc. Natl. Acad. Sci. 102: 9679-9684.

Pastalkova E, Serrano P, Pinkhasova D, Wallace E, Fenton AA, Sacktor TC (2006) Storage of spatial information by the maintenance mechanism of LTP. Science 313: 1141-1144.

Petersen CC, Malenka RC, Nicoll RA, Hopfield JJ (1998) All-or-none potentiation at CA3-CA1 synapses. Proc. Natl. Acad. Sci. USA 95: 4732-4737.

Sayer RJ, Friedlander MJ, Redman SJ (1990) The time course and amplitude of EPSPs evoked at synapses between pairs of hippocampal CA3/CA1 neurons in the hippocampal slice. J. Neurosci. 10: 826-836.

Si K, Choi YB, White-Grindley E, Majumdar A, Kandel ER (2010) Aplysia CPEB can form prion-like multimers in sensory neurons that contribute to long-term facilitation. Cell 140: 421-435.

Smolen P, Baxter DA, Byrne JH (2006) A model of the roles of essential kinases in the induction and expression of late long-term potentiation. Biophys. J. 90: 2760-2775.

Smolen P, Baxter DA, Byrne JH (2012) Molecular constraints on synaptic tagging and maintenance of long-term potentiation: a predictive model. PLoS Comput. Biol. 8: e1002620.

Song S, Sjostrom PJ, Reigl M, Nelson S, Chklovskii DB (2005) Highly non-random features of synaptic connectivity in local cortical circuits. PLoS Biol 3: e68.

Takahashi N, Kitamura K, Matsuo N, Mayford M, Kano M, Matsuki N, Ikegaya Y (2012) Locally synchronized synaptic inputs. Science 335: 353-356.

Tetzlaff C, Kolodziejski C, Timme M, Tsodyks M, Worgotter F (2013) Synaptic scaling enables dynamically distinct short- and long-term memory formation. PLoS Comput. Biol. 9: e1003307.

Trachtenberg JT, Chen BE, Knott GW, Feng G, Sanes JR, Welker E, Svoboda K (2002) Long-term in vivo imaging of experience-dependent synaptic plasticity in adult cortex. Nature 420: 788-794.

Wei Y, Koulakov AA (2014) Long-term memory stabilized by noise-induced rehearsal. J. Neurosci. 34: 15804-15815.

Whitlock JR, Heynen AJ, Shuler MG, Bear MF (2006) Learning induces long-term potentiation in the hippocampus. Science 313: 1093-1097.

Winnubst J, Lohmann C (2012) Synaptic clustering during development and learning: the why, when, and how. Front. Mol. Neurosci. 5: article 70.

Yadav A, Gao YZ, Rodriguez A, Dickstein DL, Wearne SL, Luebke JI, Hof PR, Weaver CM (2012) Morphologic evidence for spatially clustered spines in apical dendrites of monkey neocortical pyramidal cells. J. Comp. Neurol. 520: 2888-2902.

Yang G, Pan F, Gan WB (2009) Stably maintained dendritic spines are associated with lifelong memories. Nature 462: 920-925.

Yasumatsu N, Matsuzaki M, Miyazaki T, Noguchi J, Kasai H (2008) Principles of long-term dynamics of dendritic spines. J. Neurosci. 28: 13592-13608.